\documentclass[12pt]{article} %changed from \documentclass
\usepackage{amssymb,latexsym}

\newcommand\Lgp{\hbox{{\rm L}\gp}}         % Loop group
\newcommand\Lgpx{\widetilde{\Lgp}}         % extended Loop group
\newcommand\Lal{\hbox{{\rm L}$\al$}}       % Loop algebra
\newcommand\Lalx{{\widetilde{\Lal}}}    % extended Loop algebra

\newcommand\gp{\hbox{\rm\bf G}}
\newcommand\csg{\hbox{\rm\bf H}}
\newcommand\agp{\hbox{\rm\bf A}}
\newcommand\bgp{\hbox{\rm\bf B}}
\newcommand\ngp{\hbox{\rm\bf N}}
\newcommand\kgp{\hbox{\rm\bf K}}

\newcommand\Kgpt{\widetilde{\kgp}}
\newcommand\kgpt{\widetilde{\kgp}}
\newcommand\ngpt{\widetilde{\ngp}}
\newcommand\bgpt{\widetilde{\bgp}}

\newcommand\agph{\hat{\agp}}
\newcommand\bgph{\hat{\bgp}}
\newcommand\kgph{\hat{\kgp}}
\newcommand\ngph{\hat{\ngp}}

\newcommand{\al}{\mathfrak{g}}
\newcommand{\aal}{\mathfrak{a}}
\newcommand{\bal}{\mathfrak{b}}
\newcommand{\hal}{\mathfrak{h}}
\newcommand{\kal}{\mathfrak{k}}
\newcommand{\nal}{\mathfrak{n}}

\newcommand\balt{\tilde{\bal}}
\newcommand\kalt{\tilde{\kal}}
\newcommand\nalt{\tilde{\nal}}

\newcommand\aalh{\hat{\aal}}
\newcommand\balh{\hat{\bal}}
\newcommand\kalh{\hat{\kal}}
\newcommand\nalh{\hat{\nal}}

\newcommand\Ad{\hbox{\rm Ad}}           % Ad
\newcommand\ad{\hbox{\rm ad}}           % ad
\newcommand\Adt{\widetilde{\Ad}}        % Ad with tilde
\newcommand\adt{\widetilde{\ad}}        % ad with tilde

\newcommand\roots{\Delta}               % roots
\newcommand\proot{{\roots_{+}}}         % positive roots
\newcommand\sproot{\Pi}                 % simple roots

\newcommand{\kf}[2]{{\rm B}( #1, #2)}   % Killing form

\newcommand\real{\mathbb{R}}              % real #'s
\newcommand\cplx{\mathbb{C}}            % complex #'s
\newcommand\intgr{\mathbb{Z}}             % integers
\newcommand\ratnl{\mathbb{Q}}                   % rationals

\newcommand{\ddto}{{d\over dt}\big\vert_{t=0}}
\newcommand{\slnc}{sl(n,\cplx)}
\newcommand\grad{\nabla}      % gradient symbol
\newcommand\Id{\hbox{\rm Id}}

\newcommand\bipp[2]{\left\langle  \, #1,#2 \, \right\rangle }

\newcommand{\Real}{\hbox{\rm{Re}}}

\newtheorem{proposition}{Proposition}[section]
\newtheorem{lemma}{Lemma}[section]
\newtheorem{corollary}{Corollary}[section]

\newtheorem{defn}{Definition}[section]

\newenvironment{pf}{\noindent{Proof}: }{ $_\Box$ \bigskip}

\begin{document}

\title{Loop algebras, gauge invariants and a new completely integrable system}
\author{ M. Quinn \and S. F. Singer}
\maketitle

\section{Introduction}
% intro.tex
One fruitful motivating principle of much research on the family of
integrable systems known as ``Toda lattices'' has been the
heuristic assumption that the periodic Toda lattice in an affine
Lie algebra is directly analogous to the nonperiodic Toda lattice
in a finite-dimensional Lie algebra.  This paper shows that
the analogy is not perfect.  A discrepancy arises because the natural
generalization of the
structure theory of finite-dimensional simple Lie algebras is not the
structure theory of loop algebras but the structure
theory of affine Kac-Moody algebras.  In this paper we use
this natural generalization to construct the natural analog of the
nonperiodic Toda lattice.
Surprisingly, the result is not the periodic Toda lattice but a new
completely integrable system on the periodic Toda lattice phase space.
This integrable system is prescribed purely in terms of Lie-theoretic
data.
The commuting functions are precisely the gauge-invariant functions one
obtains by viewing  elements of the loop algebra as connections on a bundle
over $S^1$.

Toda lattice models belong to a general class of
integrable systems associated to vector space splittings of Lie
algebras. Suppose that a Lie algebra $\al$ (with corresponding connected
Lie group $\gp$) splits as a vector space into
$\al = \kal +\bal$, where $\kal$ and $\bal$ are subalgebras
(corresponding to subgroups $\kgp$ and $\bgp$ respectively).
One can naturally identify $\al^* \cong \kal^* + \bal^*$, and hence
coadjoint orbits of $\bgp$ can be thought of as sitting inside $\al^*$.
The invariant functions on $\al^*$, when restricted to a coadjoint orbit of $\bgp$, give a family of Poisson commuting functions
on that orbit and generate flows there described by Lax pair equations;
this is the content
of the Kostant-Symes involution theorem \cite{Ko},\cite{Sy}.
And so provided the dimensions are right one can generate interesting completely
integrable systems on the coadjoint orbits of $\bgp$.  The classical
Toda lattices arise in this way from various splittings of $sl(n,\real)$
and $\slnc$; their construction generalizes to arbitrary simple
Lie algebras (see for instance \cite{Pe}).

In the present paper the algebra considered is a central extension
$\Lalx$ of an infinite dimensional loop algebra associated to
a finite--dimensional simple Lie algebra $\al$.
In section~\ref{back} we review a well-known
construction (e.g. \cite{GW2}, \cite{RS}, \cite{RSF}) of the real symmetric 
periodic Toda phase space as a coadjoint orbit,
obtained from a splitting of $\Lalx$ in the manner described above.  
Because of the remarkable fact that coadjoint 
orbits in $(\Lalx )^*$ are parameterized by 
conjugacy classes in the finite dimensional group $\gp$,
a natural choice of invariant functions on $(\Lalx)^*$ is available via the 
class functions of $\gp$. We give a construction of these invariant functions 
in section~\ref{fns}, 
and in section~\ref{loop-reg} address a technical issue that ensures the 
generic non-degeneracy of these functions on the Toda phase space. In 
section~\ref{flows}
we show that this family of functions generates a
completely integrable system on our phase space
and that the Hamiltonian of the periodic Toda lattice is not contained in this 
family.  Finally, we have included a glossary of notation at the back.

The idea of using conjugation--invariant functions on $\gp$ to construct a 
family of commuting functions, the proof of the existence of loop--regular
elements in the Toda phase space and other results in sections \ref{fns} and 
\ref{loop-reg} were first presented in Quinn's thesis \cite{Qu}.  
The authors wish to thank Victor Guillemin, David Vogan and Allen Knutson for 
helpful conversations.  
Singer gratefully acknowledges
the support of the Bunting Institute of Radcliffe College and ONR grant 
\#N00014-89-J-3112.

\section{Background}
\label{back}
In this section we fix some notation and review the definition of
the Toda phase space.  The most difficult technical point is the
construction (due to Goodman and Wallach \cite{GW}) of an extended loop algebra
that has an associated group, as it is not true that every
infinite--dimensional Lie algebra has an associated Lie group.

First we recall a few standard facts from
Lie theory.
%, (see for example \cite{Kn}).
Take $\al$ to be a simple finite dimensional
Lie algebra over $\cplx$ of rank $\ell$,
with corresponding connected and simply connected
Lie group $\gp$.
Let $\kf{\cdot}{\cdot}$ denote the Killing form on $\al$.
Fix a Cartan subalgebra $\hal\subset\al$,
with corresponding Cartan subgroup $\csg$, 
and let $\roots$ denote
the set of roots associated to the pair
$(\al,\hal)$.

Pick a set $\sproot = \{\alpha_1,\dots,\alpha_\ell\}$
of simple roots
and let $\proot$
denote the corresponding positive roots.
Let $\alpha_*$ denote the highest root
so that $\alpha_*=\sum k_i\alpha_i$
for some positive integers $k_i$.
Choose a Chevalley basis
$\{ E_\alpha \}_{\alpha\in\roots} \cup \{ H_{\alpha_i}\}_{{{\alpha}_i}
                \in\sproot}$
for $\al$, so that $E_\alpha$
belongs to the $\alpha$ rootspace,
and $H_\alpha = [E_\alpha,E_{-\alpha}]$
is normalised by $\alpha (H_\alpha ) =2$.
Then the real span of
$$
\{ iH_\alpha, \;E_\alpha-E_{-\alpha},
\;i(E_\alpha+E_{-\alpha}): \quad \alpha\in\roots\} \;
$$
gives a compact real form $\kal$ of $\al$.
Let $\kgp$ be the corresponding compact real subgroup of $\gp$.
Then one has Cartan decompositions $\al=\kal+i\kal$ and
$\gp=\kgp\exp (i\kal)$.
Let $\aal$ be the real span of
$\{H_\alpha \}_{\alpha\in\roots}$,
and $\nal$ the complex span of
$\{E_\alpha \}_{\alpha\in{\proot}}$.
Take $\agp$ and $\ngp$ to be the corresponding subgroups
of $\gp$. Then one has Iwasawa decompositions 
$\al=\kal+\aal+\nal$ and $\gp = \kgp\agp\ngp$.
We follow convention in letting $\bgp$ denote the group
$\agp\ngp$ and letting $\bal$ denote its Lie algebra.

%%%%%%%%%%%%%%%%%%%%%%%%%%%%%%%%%%%%%%%%%%%%%%%%%%
As an example consider
$sl(n,\cplx) = su(n)+{\aal}_n +{\nal}_n$,
where $\aal_n$ denotes the algebra of real diagonal
traceless $n\times n$ matrices,
and $\nal_n$ is the algebra of strictly
upper-triangular $n\times n$ matrices.
The corresponding group factorization is
$SL(n,\cplx) = SU(n)\agp_n \ngp_n$,
where $\agp_n$ is the group of
diagonal  $n\times n$ matrices of determinant one
with real positive entries,
and $\ngp_n$ is the unipotent group
of upper-triangular matrices with 1's on the diagonal.
It follows that $\bal_\ell$ is the group of upper triangular matrices of 
determinant 1.

In general $\gp$ can be realized as a linear subgroup
of $SL(n,\cplx)$ for some $n$, in such a way that if $g\in\gp$
has $g=k(g)a(g)n(g)$ as its Iwasawa decomposition in $SL(n,\cplx)$,
then each factor belongs to $\gp$
and this factorization corresponds to the
Iwasawa decomposition in $\gp$.
We fix such a linear embedding of $\gp$, and identify
$\al$ with the corresponding Lie subalgebra of $sl(n,\cplx)$.

\bigskip

We describe now some
specific infinite dimensional groups and algebras
for which there are decompositions directly analogous
to the ones described above
(the construction is taken from \cite{GW}).
Suppose that $w$ is a symmetric weight function on $\intgr$,
i.e. $w$ is a positive function on the integers
such that $w(k+m) \leq w(k)w(m)$, and $w(k)=w(-k)$.
In addition suppose that $w$ is of non-analytic type:
$\lim_{k \to \infty} w(k)^{1/k} = 1$,
and is rapidly increasing at infinity:
$\lim_{k \to \infty} |k|^{-1/\lambda} \log w(k)= \infty$
for some $\lambda\in (1,2)$.
(For example take $w(k) = \exp (|k|^{2\over 3})$).
Take $A_w$ to be the space of functions
$f(z) = \sum_k a_k z^k$
on $S^1$ satisfying
$$
\| f \|_w := \sum_{k\in \intgr} |a_k| w(k) \; < \infty,
$$
and let $M_n(A_w)$ denote the Banach algebra of
$n\times n$ matrices
with entries in $A_w$, with norm given by
$\| T \|_w := \{ \sum_{i,j} \| T_{i,j} \|_w^2 \}^{1\over 2}$.
Define
\begin{eqnarray*}
SL_n(A_w) &=& \{ g\in M_n(A_w) \colon \; det(g)=1 \}, \\
SU_n(A_w) &=& \{ g\in SL_n(A_w) \colon \; g(z)\in SU(n) \;\;
\forall z\in S^1 \}, \\
\ngp_n(A_w) &=& \{ g\in SL_n(A_w) \colon \; g(z)=\sum_{k\geq 0}a_k
z^k,\; a_0\in \ngp \} \; .
\end{eqnarray*}
Each of these sets is a Banach-Lie group, with corresponding Banach-Lie algebra
\begin{eqnarray*}
sl_n(A_w) &=& \{ x\in M_n(A_w) \colon \; tr(x)=0 \}, \\
su_n(A_w) &=& \{ x\in sl_n(A_w) \colon \; x(z)\in su(n) \;\;
\forall z\in S^1 \},\\
\nal_n(A_w) &=& \{ x\in sl_n(A_w) \colon \;
x(z)=\sum_{k\geq 0}a_k z^k,\; a_0\in \nal  \}.
\end{eqnarray*}
One has Cartan and Iwasawa decompositions analogous to the finite
dimensional case:
\begin{eqnarray*}
sl_n(A_w)&=&su_n(A_w) + i\,su_n(A_w) = su_n(A_w) + \aal_n +\nal_n(A_w) \\
SL_n(A_w)&=&SU_n(A_w)\exp(i\,su_n(A_w)) = SU_n(A_w) \agp_n \ngp_n(A_w)
\end{eqnarray*}
Now define
\begin{eqnarray*}
\Lal &=& \{ x\in sl_n(A_w) \colon \; x(z)\in \al \;\; \forall z\in S^1 \},\\
\Lgp &=& \{ g\in SL_n(A_w) \colon \; g(z)\in \gp \;\; \forall z\in S^1 \}.
\end{eqnarray*}
\noindent
$\Lgp$ is a complex Lie subgroup of $SL_n(A_w)$
with Banach-Lie algebra $\Lal$.
%(which is simply the completion of the polynomial
%algebra $\al [z,z^{-1}]$ with respect to the norm
%$\| \cdot \|_w$). 
The algebra and group decompose as:
\begin{eqnarray*}
\Lal &=& \kalt + i\kalt =\kalt + \aal + \nalt \\
\Lgp &=& \kgpt\exp(i\kalt) =\kgpt\agp\ngpt 
\end{eqnarray*}
where $\kalt = \Lal\cap su_n(A_w)$, $\nalt = \Lal\cap \nal_n(A_w)$,
$\kgpt =\Lgp\cap SU_n(A_w)$ and $\ngpt =\Lgp\cap \ngp_n(A_w)$.
Define $\balt = \aal + \nalt$ and
$\bgpt =\agp\ngpt$ so that we have
$$
\Lal = \kalt + \balt,
\quad
\Lgp = \kgpt\bgpt.
$$
Note that elements of $\bgpt$ are analytic functions on the open unit
disk whose power series converges absolutely to a smooth function on the
unit circle.

\bigskip
From the Killing form $\kf{\cdot}{\cdot}$
on $\al$ one obtains a non-degenerate bilinear form
on $\Lal$ by integration over $S^1$, which
by an abuse of notation we also
denote $\kf{\cdot}{\cdot}$:
$$
\kf{\phi}{\psi} =
{1\over 2\pi}\int_0^{2\pi} \kf{\phi(e^{i\theta})}{\psi(e^{i\theta})}\, d\theta
$$
Let $D=z{d\over dz}$. This derivation gives a natural
2-cocycle $\omega$ on $\Lal$ defined by
$\omega (\phi, \psi) = \kf{D\phi}{\psi}$.
This  $\cplx$-valued skew-symmetric bilinear form  
can be used to construct
a central extension of $\Lal$, which we denote $\Lalx$.

Goodman and Wallach \cite{GW} have shown that
%Theorem 6.3 from that paper
provided the weight function
$w$ is of non-analytic type and is
rapidly increasing at infinity, 
there exists a Banach-Lie group whose Lie algebra is $\Lalx$.
This group is a central extension of $\Lgp$ by $\cplx^{\times}$,
and we denote it $\Lgpx$. One also has
Cartan and Iwasawa decompositions:
\begin{eqnarray*}
\Lalx &=& \kalh + i\kalh = \kalh + \aalh + \nalh \\
\Lgpx &=& \kgph \exp (i\kalh) = \kgph \agph \ngph 
\end{eqnarray*}
$\kgph$ is a
central extension of $\kgpt$ by $\rm{S}^1$,
$\agph = \agp \real^+$, and $\ngph \cong \ngpt$,
(correspondingly $\kalh =\kalt+i\real$, $\aalh=\aal+\real$,
and $\nalh\cong\nalt$).

Set $\balh = \aalh + \nalh$ and $\bgph = \agph\ngph$.
We shall apply the Kostant-Symes approach to the
splitting $\Lalx = \kalh + \balh$, and construct the Toda phase space as a coadjoint orbit of $\bgph$.

\bigskip
The next step is to calculate the coadjoint action of $\Lgpx$
on $\Lalx$; we follow the treatment in \cite{PS}.
Because $\Lgpx$ is a central extension its coadjoint action comes from an action of $\Lgp$.
The adjoint action of $\Lal$ on $\Lalx$ is
$$
\adt_{\phi}\big( \psi,q \big) = \big( \ad_{\phi} \psi, \omega(\phi,\psi) \big),
$$
and this in turn comes from an adjoint 
action of the group $\Lgp$ on $\Lalx$. 
If $\gamma \in \Lgp$ and $(\psi,q) \in \Lalx$ one has
$$
\Adt_\gamma \big( \psi,q \big) =
        \big( \Ad_\gamma \psi, q + \kf{\gamma^{-1} D\gamma}{\psi} \big).
$$
From now on we shall treat $\Lalx = \Lal + \cplx$
as an algebra over $\real$.
We introduce a nondegenerate real-bilinear pairing
$\bipp{\cdot}{\cdot}$
on $\Lalx$ defined by
$$
\bipp{(\phi,p)}{(\psi,q)}=
 {\rm Re}\big\{ \kf{\phi}{\psi} + pq \big\}.
$$
Using this form one can embed
$\Lalx \subset (\Lalx)^*$ as a real subspace,
and the image of this embedding is invariant under the
co-adjoint action of $\Lgp$.
In fact for $\gamma \in \Lgp$ and
$(\phi,p)\in \Lalx \subset (\Lalx)^*$,
the co-adjoint action of $\gamma$ on $(\phi,p)$ is given by
$$
\Adt_{\gamma}^* \big( \phi,p \big) =
        \big( \Ad_\gamma\phi + p{\gamma D\gamma^{-1}}, p \big).
$$
This action is sometimes called  the {\it gauge action} of $\gamma\in \Lgp$ on $(\phi,p)\in\Lalx$.
If two elements of $\Lalx$ are conjugate under this action
we shall refer to them as being {\it gauge-conjugate}.

\bigskip
Finally, we are in a position to construct the (symmetric, real, periodic,
tridiagonal) Toda phase space as a coadjoint orbit of the group $\bgph$.
From the splitting
$\Lalx = \kalh + \balh$ it follows that
$\Lalx^* = \kalh^* + \balh^*$.
%, by identifying
%$\kalh^*$ and $\balh^*$ with the annihilator
%of $\balh$ and $\kalh$ respectively.
Using the form $\bipp{\cdot}{\cdot}$ one has
$\kalh^\perp \subset {\rm Ann}(\kalh)\cong\balh^*$,
and this subspace is invariant under
the co-adjoint action of $\bgph$
for if $\gamma\in\bgph$ and $(\phi,p)\in\kalh^\perp$
the co-adjoint action of $\gamma$ on $(\phi,p)$
is given by
$$
\Pi_{\kalh^\perp}\; \Adt_\gamma^* (\phi,p).
$$
Here $\Pi_{\kalh^\perp}$ denotes projection 
onto $\kalh^\perp$ along $\balh^\perp$,
(one can easily verify that $\Lalx = \kalh^\perp + \balh^\perp$
where $\kalh^\perp = i\kalh$ and $\balh^\perp = i\balh$).

The symmetric periodic Toda phase space is defined to be the
orbit of $\bgph$ through the element
$(\phi_o,1)\in \kalh^\perp$
where
$$
\phi_o(z)=\sum_i (E_{\alpha_i}+E_{-\alpha_i})
+ E_{\alpha_*}z^{-1} + E_{-\alpha_*}z,
$$
(recall that $\alpha_*=\sum k_i\alpha_i$ is the
highest root of $\al$).
This orbit consists of all elements of the form $(\phi,1)$ with
$$
\phi = H+\sum_i q_i(E_{\alpha_i}+E_{-\alpha_i})
+ q_*(E_{\alpha_*}z^{-1} + E_{-\alpha_*}z),
$$
where $q_i,q_*>0$ satisfy
$q_* \cdot \prod  q_i^{k_i} =1$,
and $H$ belongs to $\aal$.
The periodic Toda Hamiltonian is 
${\rm H} (\phi) = {1\over 2}\kf{\phi}{\phi}$.

\bigskip
%*** review Factorization Theorem
We shall apply the involution theorem of Kostant and Symes and the factorization
theorem of Adler and van Moerbeke and Reyman and Semenov--Tian--Shansky in this infinite dimensional setting.  For the details of these theorems, see \cite{RS}.
Because $\Lalx=\kalh+\balh$ is a splitting of $\Lalx$ as a vector sum of
subalgebras and because $(0,1)\in \Lalx^{*}$ satisfies
$\langle(0,1),[\kalh,\kalh]\rangle=0$ and $\langle(0,1),[\balh,\balh]\rangle=0$ we can apply
the involution theorem to show that the restrictions of any coadjoint-invariant functions on $\Lalx$ to the Toda phase space will commute in the
standard Poisson structure.  Furthermore, the factorization theorem tells
us that if $I$ is a coadjoint--invariant function then the Hamiltonian flow
associated to $I$ on the Toda phase space is given by
$$(\phi(t),1)=\Adt^{*}_{b(t)}(\phi(0),1),$$
where exp $t \nabla I(\phi(0),1)=k(t)^{-1}b(t)$ with $k(t)\in \kgph$ and
$b(t)\in \bgph.$

\section{Loop--regularity and invariant functions}
\label{fns}
We shall construct a family of
Poisson commuting functions on the Toda phase space
by an application of the
Kostant-Symes involution theorem.
To do so we require invariant functions on $(\Lalx)^*$, (or,
more precisely, on $\Lalx\subset (\Lalx)^*$).
Roughly speaking, coadjoint orbits in
$\Lalx\subset (\Lalx)^*$ 
correspond to conjugacy classes in $\gp$, and so a
natural way to get invariant functions on $\Lalx$ is 
via the class functions of $\gp$. 
We make this precise as follows.

\begin{defn}
Given $(\phi,p)\in\Lalx$ such that $p\neq 0$
the differential equation
$$
{\dot f(t)}f(t)^{-1}={1\over ip}\phi(e^{it}), \quad f(0)=\Id,
$$
where $f\colon\real\to\gp$
is called the {\rm monodromy equation} 
associated to $(\phi,p)$.
The {\rm monodromy} ${\rm M}_{(\phi,p)}$ 
of the element $(\phi,p)\in\Lalx$ is
defined by ${\rm M}_{(\phi,p)}=f(2\pi)$, and we call
$$
{\rm M}\;\colon \Lalx\rightarrow\gp \;\colon (\phi,p)\mapsto {\rm M}_{(\phi,p)}
$$ 
the {\rm monodromy map}.
An element $(\phi,p)\in\Lalx$ will be said to be 
{\rm loop-regular} if its monodromy is regular, 
i.e. if the normalizer of 
${\rm M}_{(\phi,p)}$ in $\gp$ is conjugate to the Cartan subgroup $\csg$.
\end{defn}
Since $\Lalx$ is a Banach space, by standard results from the 
theory of differential equations the monodromy map is smooth.
The loop-regular elements form an open subset of $\Lalx$,
and in section \ref{loop-reg} we show that they are dense in the Toda phase space.

The following proposition from \cite{PS}
gives a parametrization of the co-adjoint orbits
and allows one to describe the invariant functions on 
$\Lalx \subset (\Lalx )^*$.

\medskip

\begin{proposition}\cite{PS}
\label{press-seg}
The monodromy classifies the gauge co-adjoint orbits in the following
sense{\rm{:}}
\begin{enumerate}
\item If $\gamma\in\Lgp$, then the monodromy of 
$\Adt^*_\gamma(\phi,p)$ is 
$\gamma(1){\rm M}_{(\phi,p)}\gamma(1)^{-1}$.

\item If ${\rm M}_{(\phi,p)}$ and ${\rm M}_{(\tilde{\phi},p)}$ 
are conjugate in $\gp$, then $(\phi,p)$ and $(\tilde{\phi},p)$
are gauge-conjugate in $\Lalx$. 
Specifically, if 
${\rm M}_{(\phi,p)} = g{\rm M}_{(\tilde{\phi},p)}g^{-1}$ 
for some $g\in\gp$, and if $f$ and $\tilde{f}$ solve the
monodromy equation for $(\phi,p)$ and $(\tilde{\phi},p)$
respectively, then $(\phi,p) = \Adt^*_\gamma (\tilde{\phi},p)$
where $\gamma(e^{it}) = f(t)g\tilde{f}(t)^{-1}$.
\end{enumerate}
\noindent
Hence for fixed $p\neq 0$, the monodromy map gives a 1-1 correspondence
between coadjoint orbits of $\Lgp$ in
$L\al\times\{p\}\in\Lalx$ and conjugacy classes of $\gp$.
\end{proposition}

A simple corollary to this proposition is that loop-regular elements can be conjugated to
$\hal\times\cplx^\times$ under the gauge action:
\begin{corollary}
\label{invnt1}
For any loop-regular $(\phi,p)$ there exists $\gamma\in\Lgp$ and a
constant loop $\mu\in\hal\subset\Lal$ such that
$(\phi,p)=\Adt^*_\gamma(\mu,p)$. \end{corollary}
\begin{pf}
By loop-regularity ${\rm M}_{(\phi,p)}$ 
is conjugate to some $x\in\csg$.
Choose $\xi\in\hal$ such that $x=\exp\xi$
and set $\mu ={ip\over 2\pi}\xi$.
By construction ${\rm M}_{(\mu,p)}=x$, 
so that ${\rm M}_{(\phi,p)}$ and
${\rm M}_{(\mu,p)}$ are conjugate in $\gp$.
Hence by the second part of proposition \ref{press-seg},
$(\phi,p)$ and $(\mu,p)$ are gauge-conjugate.
i.e. there exists $\gamma\in\Lgp$ with 
$(\phi,p)=\Adt^*_\gamma(\mu,p)$.
\end{pf}

In the case of loop-regular elements in the Toda
phase space the same argument can be strengthened to
the following:

\begin{corollary}\cite{Qu}
\label{todconj}
For any loop-regular $(\phi,1)$ in the Toda phase space
there exists $\gamma \in \kgpt \subset \Lgpx$,
and $\mu \in\aal\subset\hal$, such that
$(\phi,1) = \Adt^*_\gamma(\mu,1)$.
\end{corollary}

\begin{pf}
The monodromy equation for $(\phi,1)$ is
$\dot{f}(t)f^{-1}(t) = {1\over i}\phi (e^{it})$.
Since $\phi(z)\in i\kalt$, this shows that $f(t)\in\kgp$ for all $t$, 
and in particular the monodromy $f(2\pi) = {\rm M}_{(\phi,1)}$ is in $\kgp$. 
By loop-regularity, and since $\kgp$ is the compact form of $\gp$, 
${\rm M}_{(\phi,1)}$ can be conjugated in $\kgp$ to an element of 
the maximal torus $\csg \cap \kgp$ of $\kgp$.
Thus there exists $k\in\kgp$ and $x\in\csg\cap\kgp$
with ${\rm M}_{(\phi,1)}=kxk^{-1}$.

Choose $\xi\in\hal\cap\kal =i\aal$ such that $x=\exp\xi$
and set $\mu={i\over 2\pi}\xi$.
By construction $\tilde{f}(t) = \exp {t\over 2\pi}\xi$
solves the monodromy equation for $(\mu,1)$,
and ${\rm M}_{(\mu,1)}=x$ is conjugate to ${\rm M}_{(\phi,1)}$.
Hence $(\phi,1)$ and $(\mu,1)$ are gauge-conjugate
by proposition \ref{press-seg}, and
specifically one has $(\phi,1)=\Adt^*_\gamma{(\mu,1)}$
where $\gamma(e^{it}) = f(t)k\tilde{f}(t)^{-1}$.
By construction $\mu\in\aal$, and the loop $\gamma$
has image in $\kgp$, so $\gamma\in\kgpt$ as required.
\end{pf}

%%%%%%%%%%%%%%%%%%%%%
\bigskip
We now define a family of invariant functions on $\Lalx$,
which when restricted to the Toda phase space will give rise
to a completely integrable system. We must first define a factorization.
Given $g\in\gp$ let ${\rm k}[g]$ denote the
$\kgp$ factor of $g$ in the Cartan decomposition
$\gp=\kgp\exp(i\kal)$.
Next, let $\Psi_1,\dots,\Psi_{\ell}$ denote the characters of the irreducible 
representations corresponding to the fundamental weights of $\gp$. Then 
$d\Psi_1,\dots,d\Psi_{\ell}$ are independent at regular values of $\gp$ 
\cite[pg. 123]{St}, and because the fundamental weights occur in dual pairs the characters occur in conjugate pairs. Thus their real and imaginary  parts give a collection of $\ell$ r
eal-valued class functions on $\gp$ which are also functionally independent 
at regular values of $\gp$. We denote this collection 
$\chi_1,\dots,\chi_{\ell}$.
\begin{defn}
\label{fndefn}
For $j=1,\dots,\ell$, define
$$
F_j \;\colon \Lalx \to \real \;\colon \;(\phi,p) \mapsto \chi_j\left( {\rm k}
		\!\!\left[ {\rm M}_{(\mu,p)}\right] \right)
$$
where $(\mu,p)$ is any loop in $\hal\times\cplx$ gauge-conjugate to $(\phi,p)$.
\end{defn}

For 
$\mu\in\hal$ one has ${\rm M}_{(\mu,p)} = \exp ({2\pi \over i}{\mu\over p})$, 
and ${\rm k}\!\!\left[ {\rm M}_{(\mu,p)} \right] = \exp \left({2\pi \over i}
		\Pi_{\aal}{\mu\over p}\right)$,
so that an equivalent expression for $F_j$ is
$$
F_j(\phi,p) = \chi_j \left( \exp \left({2\pi \over i}\Pi_{\aal}{\mu\over p}
		\right) \right)\; ,
$$
whenever $(\phi,p)$ is gauge-conjugate to $(\mu,p)$.
To see that $F_j$ is well-defined
suppose that $(\mu,p)$ and $(\tilde{\mu},p)$ are gauge-conjugate elements of $\hal\times\cplx$. Then they are conjugate by an element of the affine Weyl group so 
$\tilde{\mu} = w\cdot\mu + p\lambda$
for some $w$ in the ordinary Weyl group, and $\lambda$ in the coroot lattice.
Since $\exp (2\pi i\lambda) = \Id$ and the Weyl group 
preserves $\aal$ it follows that
$$
\exp \left({2\pi \over i}\Pi_{\aal}{\tilde{\mu}\over p}\right) = \exp \left(w\cdot{2\pi \over i}\Pi_{\aal}{\mu\over p}\right) =
k\exp \left({2\pi \over i}\Pi_{\aal}{\mu\over p}\right)k^{-1} \; 
$$
for some $k\in\kgp$.
Hence the functions $F_j$ are well-defined and invariant under the gauge action.   

%%%%%%%%%%%%%%%%%%%%%%%%%%%%%%%%%
\medskip
In order to analyze the flows induced by the functions $F_j$ 
on the Toda phase space we introduce a family of locally 
defined functions that are simpler to work with.
To construct these local functions we first show that gauge-conjugation to $\hal\times\cplx^\times$
is (locally) a well defined map.
\begin{proposition} \label{prop:eta}
Fix any loop-regular $(\psi,q)$ in $\Lalx$ 
and any $(\xi,q)\in\hal\times\cplx^\times$ gauge-conjugate to $(\psi,q)$. 
There exists a gauge-invariant loop-regular neighborhood $U\subset\Lalx$ of $(\psi,q)$, and a smooth gauge-invariant map 
$$
\eta \colon U\to\hal\times\cplx^\times
$$ 
such that $\eta(\psi,q)=(\xi ,q)$,
and $\eta(\phi,p)$ is gauge-conjugate to $(\phi,p)$ for each $(\phi,p)\in U$.
\end{proposition}

\begin{pf}
This proposition follows from the implicit function theorem for Banach 
manifolds.
%, although in this setting we need the Nash-Moser inverse %function theorem for tame Fr\'echet spaces, (see \cite{Ha} for %more details).
We sketch the argument.

Let $\Psi =(\Psi_1,\dots,\Psi_{\ell})$ 
denote the vector of character maps
corresponding to the fundamental weights of $\al$. 
Define 
$F\colon\Lalx\times(\hal\times\cplx^\times)\to \cplx^l\times\cplx$
by
$$ 
\left( (\phi,p),(\mu,r) \right) \mapsto 
\left( \Psi( {\rm{M}_{(\phi,p)}} ) -  \Psi( {\rm{M}_{(\mu,r)}} ), p-r \right) \; .
$$
Then by hypothesis $F((\psi,q), (\xi,q)) = (0,0)$, and one can easily check that ${\rm{D}_2} F((\psi,q), (\xi,q))$ is invertible, (the loop-regularity of $(\xi,q)$ is needed here).
So by the implicit function theorem there is a neighborhood $\tilde{U}$ of 
$(\psi,q)$ and a smooth map 
$\tilde{\eta} \colon \tilde{U}\mapsto\hal\times\cplx^\times$ 
such that
$F((\phi,p), \tilde{\eta}(\phi,p)) = (0,0)$, and
$\tilde{\eta}(\psi,q) = (\xi,q)$.  
Since $(\xi,q)$ is loop-regular 
its stabiliser in the affine Weyl group
is trivial, and there is a neighborhood 
of $(\xi,q)$ in $\hal\times\cplx^\times$
 on which no two elements are gauge-conjugate. 
Shrinking $\tilde U$ if necessary we can 
assume that $\tilde\eta(\tilde U)$ lies in such a neighborhood. 
Now set $U=\Adt^*_{\Lgp}\tilde U$ 
and define $\eta\colon U\to\hal\times\cplx^\times$ to be the map which 
first gauge-conjugates
$(\phi,p)\in U$ to $\tilde U$, then maps to 
$\hal\times\cplx^\times$ by $\tilde\eta$. 
By construction $\eta$ has
the desired properties.
(Note that since the set of loop-regular elements is open in $\Lalx$, we can without loss of generality take all the open sets constructed to be loop-regular).
\end{pf}

Quinn has proved a stronger version of this result in his thesis:
the map $\eta$ can be extended smoothly to the central extension of the 
algebra of smooth loops in $\al$ \cite{Qu}.  In that case the theory of
 Banach manifolds does not apply, and one needs the Nash-Moser implicit
function theorem.

Let $\hat\eta$ denote the composition of projection onto
$\hal$ with $\eta$.  (For example, in the proposition above
$\hat\eta (\psi,q) = \xi$.)
We now define a family of gauge-invariant 
functions on the set $U$ constructed in proposition~\ref{prop:eta}.
\begin{defn}
For $j=1,\dots,\ell$, define 
$$
I_j \; \colon U\to \real \; \colon(\phi,p)\mapsto 
\Real \left[ \alpha_j ({\hat\eta(\phi,p)\over p})\right]
\;=\alpha_j \left( \Pi_{\aal}{\hat\eta(\phi,p)\over p}\right) .
$$
\end{defn}
Take $\tau^j \in\aal$, $j=1,\dots,\ell$, to be a dual
basis to the simple roots, so that 
$\alpha_i (\tau^j ) = \delta_{ij}$.
By construction
$\sum_k \tau^k I_k (\phi,p) = \Pi_{\aal} {\hat{\eta}(\phi,p)\over p}$,
and so 
$$
F_j |_U  =
\chi_j \left(\exp\left( {2\pi\over i} \sum_k \tau^k I_k \right) \right)\; .
$$
While this characterization of the $F_j$'s holds only locally, we use it
to show functional independence of the family.

\bigskip
We finish this section with a calculation 
of the gradients of the functions $I_j$. 
The following lemma makes the calculation relatively easy.

\begin{lemma}
\label{vectsplit}
Suppose $(\mu,p)$ is loop-regular and $\mu\in\hal$.
Then $\Lalx$ has vector space splitting
$$
\Lalx = \left[ \hal\times\cplx \right] \oplus \adt^*_{\Lal} (\mu,p),
$$
and this splitting is orthogonal with respect to the form
$\bipp{\cdot}{\cdot}$.
\end{lemma}

\begin{pf}
To show that $\hal\times\cplx$ and $\adt^*_{\Lal} (\mu,p)$
are orthogonal is straightforward. Take any $\nu\in\Lal$
and any $(\mu',p')\in\hal\times\cplx$, then
\begin{eqnarray*}
\bipp{\tilde{\ad}_{\nu}^* (\mu,p)}{(\mu',p')}
&= & \bipp{(\mu,p)}{-\adt_{\nu} (\mu',p')} \\
&= & -\bipp{(\mu,p)}{([\nu,\mu'],0)} \; .\\
&= & -\Real \, \kf{\mu}{[\nu ,\mu']} %\\
%&= & \Real \, \kf{[\mu,\mu']}{\nu}\\
%&= & 0,
\end{eqnarray*}
%where $\kf{\cdot}{\cdot}$ denotes the bilinear form obtained by %integrating
%the Killing form (see Section \ref{back}).  The last equality %holds
The last expression is zero by invariance of the Killing form, and because both $\mu$ and $\mu'$ lie in the commutative algebra $\hal$.

To show that $\hal\times\cplx$ and $\adt^*_{\Lal} (\mu,p)$ together
span $\Lalx$ it suffices to show that given any $(\psi,q)\in\Lalx$
one can find $\xi\in\hal,\; r\in\cplx$, and $\nu\in\Lal$ such that
$$
(\psi,q) = (\xi,r) + \adt^*_{\nu} (\mu,p).
$$
Suppose $\psi = \sum \psi_k z^k$, and that
$\psi_k = \psi_k^o +\sum_{\alpha\in\roots} \psi_k^{\alpha}E_{\alpha}$
is the rootspace decomposition of $\psi_k$ in $\al$, with 
$\psi_k^o\in\hal$.
Then $r=q$, $\xi=\psi_0^o$, and 
$\nu=\sum\nu_k z^k$ with 
$\nu_k = \nu_k^o +\sum_{\alpha\in\roots} \nu_k^{\alpha}E_{\alpha}$
where
$$
\nu_k^{\alpha} = {-1\over pk +\alpha(\mu)}\psi_k^{\alpha}, \quad\quad
(k\neq 0)\;\; \nu_k^o = {-1\over pk}\psi_k^o .
$$
(Note that loop-regularity of $(\mu,p)$ ensures 
that $pk + \alpha(\mu)$ is never zero). 
$\nu_0^o\in\hal$ can be chosen arbitrarily.
Thus $[\hal\times\cplx]\oplus\adt^*_{\Lal}(\mu,p)$ does span $\Lalx$.

That $\hal\times\cplx$ and $\adt^*_{\Lal}(\mu,p)$ intersect 
trivially now follows from their orthogonality, and the 
non-degeneracy of the form $\bipp{\cdot}{\cdot}$.
Hence as required
$\Lalx = \left[ \hal\times\cplx \right] 
\oplus \adt^*_{\Lal} (\mu,p)$.
\end{pf}

For each $\alpha_j \in\Pi$ take $h_j \in\hal$ to be 
the unique element satisfying $\alpha_j (x) = \kf{x}{h_j}$ for every $x\in\hal$. (Note in particular that $h_j\in\aal$). 
The gradient of $I_j$ is now straightforward to calculate:

\begin{proposition}
\label{gradI}
Let $\eta$ and $U$ be as in proposition~\ref{prop:eta}.
Suppose $(\phi,p)\in U$, $\gamma\in\Lgp$ and
$\mu\in\hal$, are such that 
$(\phi,p)=\Adt^*_\gamma (\mu,p)$
where $\eta(\phi,p) = (\mu,p)$.
Then 
$$
\nabla I_{j}(\phi,p) =
\Adt_{\gamma}\left({1\over p}h_j,
                \; -{1\over p^2}\alpha_j (\mu )\right) \; .
$$
\end{proposition}

\begin{pf}
Because of the invariance of $I_j$ one has
$\nabla I_j(\phi,p) = \Adt_{\gamma} \nabla I_j (\mu,p)$.
Take $(\psi,q)\in\Lalx$ and decompose it as
$(\psi,q) =(\xi,q) +\adt_{\nu}^*(\mu,p)$,
as in lemma~\ref{vectsplit},
with $\xi\in\hal$ and $\nu\in\Lal$. Then
$$
\bipp{(\psi,q)}{\grad I_j (\mu,p)} 
=\bipp{(\xi,q)}{\grad I_j (\mu,p)} 
        +\bipp{\adt^*_\nu(\mu,p)}{\grad I_j (\mu,p)}.
$$
Note that the second term is zero because of the
invariance of $I_j$. The first term is 
$\ddto I_j(\mu + t\xi, p+tq)$, and since by construction
$\eta$ fixes a neighborhood of $(\mu,p)$ in $\hal\times\cplx$,
\begin{eqnarray*}
\bipp{(\xi,q)}{\grad I_j (\mu,p)}
&=& \ddto \Real\; \alpha_j \left( {\mu+t\xi \over p+tq} \right)\\
&=& \Real \left( {1\over p}\alpha_j (\xi ) - 
        q{\alpha_j (\mu)\over p^2} \right) \\
&=& \bipp{(\xi,q)}{\left({1\over p}h_j, 
                        -{1\over p^2}\alpha_j(\mu )\right)}.
\end{eqnarray*}
By the orthogonality of $\hal\times\cplx$ and 
$\adt^*_{\Lal}(\mu,p)$
the last expression is equal to 
$\bipp{(\psi,q)}{({1\over p}h_j, -{1\over p^2}\alpha_j(\mu ) )}$,
and we get 
$\nabla I_j (\mu,p)=({1\over p}h_j, -{1\over p^2}\alpha_j(\mu ) )$, which completes the proof.
\end{pf}

\bigskip

Let $(\phi,1)$ be a loop-regular element of the Toda phase space. It follows from combining the above proposition with corollary \ref{todconj} that the $\grad I_j(\phi,1)$'s are linearly independent elements of $i\kalh =i\kalt+\real$. 
Now define $f_j\colon\real^{\ell}\rightarrow\real \colon\vec{x}\mapsto\chi_j(\exp ({2\pi\over i}\sum \tau^k x_k))$, so that locally one has $F_j\vert_U = f_j(I_1,\dots,I_{\ell})$,
and $\grad F_j =\sum_k {\partial f_j\over \partial x_k} \grad I_k$. One can easily check that because of the functional independence of the $\chi_j$'s at regular elements of $\gp$ the matrix 
$\left( {\partial f_j\over \partial x_k}(\vec{I}(\phi,1))\right)$ is invertible.
Putting these pieces together we arrive at:

\begin{corollary}
\label{globalgrad}
Let $F=\sum c_j F_j$ for some real constants $c_j$.
If $(\phi,1)$ is a loop-regular element of the Toda phase space then $\grad F(\phi,1)\in i\kalh$, and 
furthermore $\grad F(\phi,1) =0$ if and only if $c_1=\cdots=c_{\ell}=0$.
\end{corollary}

\section{Existence of loop-regular elements}
\label{loop-reg}
We now show that loop-regular elements form a dense open subset of the Toda phase space.

On $\Lal\times\{ 1\}$
the monodromy can be regarded as a map
${\rm M}:\Lal\to\gp:\phi\mapsto f(2\pi)$,
where $f\colon \real \to \gp$ solves the monodromy equation
$\dot f f^{-1} = {1\over i} \phi (e^{it})$, and $f(0)= \Id$.
This map extends to the whole of 
$C^\infty (\real ,\al )$, thinking of 
$\Lal\subset C^\infty (\real ,\al )$ as the subspace of $2\pi$-periodic functions.
By an abuse of notation we shall denote this map by the same letter,
${\rm M}\colon C^\infty (\real ,\al )\to\gp$,
and refer to it as the {\it lift map}.

The reason for this nomenclature is as follows.
If we consider the trivial principal bundle 
$\gp \hookrightarrow \real\times \gp \to \real$, 
(whose space of connections can be identified with
${\Omega^1 (\real )}\otimes \al \cong {C^\infty} (\real ,\al )$ ), then the
horizontal lift of the curve $c(t)\colon t \mapsto t$ through $(1,\Id )$
determined by the connection ${1\over i}\phi$ is given by $(c(t),f(t))$
where $f(t)$ solves the lift equation for $\phi$.
This interpretation also shows that the solution $f(t)$ exists for all
$t\in\real$.

It is not hard to verify the following lemma.
\begin{lemma}
\label{monsum}
If $f$ solves the lift equation for $\phi\in C^\infty (\real ,\al )$
and if $\psi$ is an element of $C^\infty (\real ,\al )$, then
$$
{\rm M}(\phi + \psi) = {\rm M}(\phi){\rm M}(\Ad_{f^{-1}} \psi) \; .
$$
\end{lemma}
We can now prove the following proposition:
\begin{proposition}\cite{Qu}
The set of loop-regular elements is open and
dense in the Toda phase space.
\end{proposition}
\begin{pf}
Since the monodromy is a real analytic mapping on the Toda phase space 
it suffices to show that the set of loop-regular elements of the
Toda phase space is non-empty.
We do this by constructing sequences
$\phi_k, \psi_k \in {C^\infty }(\real ,\al )$
such that $\phi_k + \psi_k$ lies in the Toda phase space,
and the following properties hold:
\begin{enumerate}
\item ${\rm M}(\phi_k)={\rm M}_0$ is constant in
$k$ and ${\rm M}_0$ is regular in $\gp$.
\item The solution $f_k :\real \mapsto \gp$
to the lift equation for $\phi_k$
has image in the compact subgroup $\kgp \subset \gp$.
\item $\psi_k\in\Lal$ and $\psi_k \to 0$ as $k\to \infty$.
\end{enumerate}
Properties (2) and (3) force $\Ad_{f_k^{-1}} \psi_k \to 0$ as $k\to\infty$,
and so by continuity of ${\rm M}$ one has ${\rm M}(\Ad_{f^{-1}_k}
\psi_k)\to\Id$.
Thus ${\rm M}(\phi_k +\psi_k)\to {\rm M}_0$ by property (1) and lemma \ref{monsum}.
Since ${\rm M}_0$ is regular and the set of regular
elements in $\gp$ is open, for sufficiently large $k$
the monodromy ${\rm M}(\phi_k + \psi_k)$ will be regular and hence $\phi_k +\psi_k$ will be loop-regular.
So the problem becomes one of finding a sequence $\phi_k + \psi_k$
satisfying (1), (2) and (3).

Consider the set $\{ H_\alpha \}_{\alpha\in\roots}$.
This forms a root system in
$\hal$, with a base given by the simple co-roots
$\{ H_{\alpha_i}\}_{\alpha_i \in \sproot}$. Define
$H = \sum_{\alpha > 0} H_\alpha$,
and let $\{ \tau^i : i=1,\dots,\ell \} \subset \hal$ be a dual basis to
the set of simple roots so that $\alpha_i (\tau^j ) = \delta_{ij}$
for all $i,j$. From the theory of root systems one has that $$
H=\sum_{\alpha > 0} H_\alpha = \sum_{i=1}^{\ell} d_i H_{\alpha_i}
=2\sum_{i=1}^{\ell} \tau^i \; ,
$$
where the coefficients $d_i$ belong to $\intgr_{>0}$. Define
$$
E=\sum_{i=1}^{\ell} \sqrt{d_i} E_{\alpha_i}, \quad
F=\sum_{i=1}^{\ell} \sqrt{d_i} E_{-\alpha_i} \; . $$
Using the three expressions for $H$ it is easy to verify that
the triple $\{H,E,F\}$ generates an algebra isomorphic to
$sl(2,\cplx)$, with $[E,F]=H$, $[H,E]=2E$ and $[H,F]=-2F$.
Consequently there exists an element
$g\in\gp$ such that $E+F = \Ad_g H$.

Recalling that $\alpha_* =\sum k_i\alpha_i$ is the
highest root of $\al$ we let $c = (\sum k_i) +1$
and set $p(k)={1+2ck \over 2c}$.
We can now define the sequences $\phi_k, \psi_k$:

\begin{eqnarray*}
\phi_k
&= &p(k)(E+F)= p(k)\sum_{i=1}^{\ell} \sqrt{d_i}(E_{\alpha_i} + E_{-\alpha_i})\;,\\
\psi_k
&= &{1\over \prod_{i=1}^{\ell}
        (p(k)\sqrt{d_i})^{k_i}}(E_{\alpha_*}z^{-1} + E_{-\alpha_*}z) \; .
\end{eqnarray*}

\medskip

One can easily check that
$\phi_k +\psi_k$ belongs to the Toda phase space for each positive integer $k$,
and by construction $\psi_k \to 0$ as $k\to \infty$. The solution to the
lift equation for $\phi_k$ is $f_k(t) = \exp {t\over i}\phi_k$,
and since ${1\over i}\phi_k$ belongs to the algebra
$\kal$ it follows that $f_k(t)$ lies in $\kgp$ for all $t$.

Since $E+F=\Ad_g H$ we have $\phi_k = \Ad_g (p(k)H)$, so that
\begin{eqnarray*}
{\rm M}(\phi_k) = f_k(2\pi) = \exp (-2\pi i \phi_k)
&=& g\exp( -2\pi i {1+2ck \over 2c}H)g^{-1} \\
&=& g\exp( -2\pi i {1\over 2c} H)\exp(2\pi iH)^{-k}g^{-1} \; .
\end{eqnarray*}
But $\exp (2\pi i H_\alpha) = \Id$, (in the adjoint representation
$H_\alpha$ is diagonal with integer entries), and so $\exp (2\pi iH) =
\Id$.
One also has that for each positive root $\alpha$, $$
\alpha\left({1\over 2c}H\right) 
                    = {1\over c} \alpha \left(\Sigma \epsilon_i\right)
                = {1\over c} \hbox{\rm ht}(\alpha)\;
\in \ratnl\cap (0,1), $$
where ${\rm ht}(\alpha)$ denotes the height of $\alpha$,
so that $\exp (-2\pi i {1\over 2c}H)$ is regular in $\gp$. This implies
that we have regularity of the monodromy
$${\rm M}(\phi_k) = g\exp (-2\pi i{1\over 2c}H)g^{-1}= {\rm M}_0 \; .
$$
Thus the requirements (1), (2) and (3) are
satisfied, and the proof is complete.
\end{pf}

\section{The flows}
\label{flows}

In this section we prove that the period lattice of the
Hamiltonian flows of the $F_j$'s is trivial.
We conclude that these functions form a completely integrable system and that
this system is not the periodic Toda lattice.
\begin{lemma}
\label{steph1}
If $(\mu,1)$ is a loop-regular element with $\mu\in\hal$,
if $\gamma\in\Lgpx$ and if $\Adt^{*}_{\gamma}(\mu,1)=(\mu,1)$,
then $\gamma$ is a constant loop in $\csg$.
\end{lemma}

\begin{pf}
Define $f(t)= \exp(-i\mu t).$ Note $f(t)\in \csg$
for all $t$.
 Then $f$ solves the monodromy equation for $(\mu,1):$
\[\frac{df(t)}{dt} f(t)^{-1} = {1\over i}\mu.\]
 Since $\Adt^{*}_{\gamma}(\mu,1) = (\mu,1)$, it follows
that $\gamma(e^{it}) f(t) \gamma(1)^{-1}$ also solves the
monodromy equation for $(\mu,1)$.
So for all $t$ we have $\gamma(e^{it}) f (t) \gamma(1)^{-1}= f(t).$
By the loop-regularity of $(\mu,1), f(2\pi)$ is regular in $\csg$, so
$\gamma(1)\in \csg.$

 Now for any $t$ we have
$\gamma(e^{it}) = f(t) \gamma(1)f(t)^{-1} = \gamma(1)$, since $f(t)$
and $\gamma(1)$ both lie in $\csg.$  So $\gamma$ is a constant
loop in $\csg$.
\end{pf}

\bigskip

\begin{lemma}
\label{steph2}
Suppose $\left( \phi,1\right)$ is a loop-regular
element of the periodic
Toda lattice phase space.  Suppose $b\in\bgpt$ and
\[\Adt^{*}_{b}(\phi,1) = (\phi,1)\]
Then $b=\Id$.
\end{lemma}

\begin{pf}
First, we claim that $b\in\ngpt$.  Write $b=an$ with
$a\in\agp$ and $n\in\ngpt$.
We will show that $a=\Id$.  Let $\left( \ \ \right)_{-1}$ denote the
natural projection onto the height $-1$ subspace
of $\Lalx$.
(The algebra $\Lalx$ has a rootspace decomposition analogous to the
finite dimensional case and the notion of root height is well-defined,
 inducing a $\intgr$-grading on $\Lalx$. See for example \cite{RSF}).
Then
$$
\left( \phi,1\right) = \Adt^{*}_{a}\Adt^{*}_{n}\left(\phi,1\right)
$$
and hence
\begin{eqnarray*}
\left(\phi,1\right)_{-1}&=&\Adt^{*}_{a}\left(
\Adt^{*}_{n}\left(\phi,1\right)\right)_{-1} \\
                                &=& \Adt^{*}_{a}\left(\phi,1\right)_{-1}.
\end{eqnarray*}
Here the first equality holds because height is
invariant under $\Adt^{*}_{a}$.
Since $\ngpt$ is generated by $\exp\nalt$, and $\nalt$ is the sum of
rootspaces of positive height, we get the second equality after noting that
Toda lattice elements have no component of height less than $-1$.

However, because $\left(\phi,1\right)$ is in the Toda
lattice phase space we know that
there exist strictly positive real numbers
$q_*$ and $q_i$, $i=1,\dots,\ell$ such that
$$
\left(\phi,1\right)_{-1}=\left(\sum_{i}q_{i}E_{-\alpha_i}
      + z^{-1}q_{*}E_{\alpha_*},0\right).
$$
Because $a\in\agp$ we can write $a=\exp(X)$ for a unique $X\in\aal$. Then
$$\Adt^*_a (\phi,1)_{-1} =
                \left(\sum_{i}e^{-\alpha_i(X)}q_{i}E_{-\alpha_i}
      + z^{-1}e^{\alpha_*(X)}q_{*}E_{\alpha_*},0\right)
$$
Thus for each $\alpha\in\sproot$ we have
$e^{\alpha(X)}=1$.
Hence $X=0$ and $a=\Id$, so that $b\in\ngpt$.
It follows that the constant term in the power series expansion of $b$,
(which is just $b(0)$), lies in $\ngp$.
\bigskip

Since $(\phi,1)$ is loop-regular, there is a $\mu\in\hal$ and a
$\gamma\in\Kgpt$ such that
\noindent $\left(\phi,1\right)=
\Adt^{*}_{\gamma}\left(\mu,1\right)$ by corollary \ref{todconj}.
Hence
$$
\Adt^{*}_{\gamma^{-1}b\gamma}\left(\mu,1\right)=\left(\mu,1\right),
$$
which implies that $\gamma^{-1}b\gamma\in\csg$ by lemma \ref{steph1}.
Write $\eta=\gamma^{-1}b\gamma$.
Then for any $z \in S^{1},$ we have $b \left( z \right)=\gamma \left( z
\right) \eta \gamma \left( z \right)^{-1}$.

Using the realization of $\gp$ as a linear subgroup of $SL(n,\cplx)$
described in section \ref{back},
now consider the characteristic polynomial
$\det\left(b(z)-\lambda\right)$.
Since $b(z)\in\bgpt$ this is an analytic function of $z$ on the open
unit disk, which is constant on the boundary of the disk
since $\det\left(b(z)-\lambda\right)=\det\left(\eta-\lambda\right)$
for any $z\in S^1$. By an application of the maximum modulus principle
it follows that this equality actually holds throughout the interior
of the unit disk as well. In particular setting $z=0$ and using the fact
that $b(0)$ lies in $\ngp$ gives
$$
\det\left(\eta-\lambda\right)=\det\left(b(0)-\lambda\right)
=\left(1-\lambda\right)^n \; .
$$
So $\eta=\Id$ and
$b(z) = \gamma (z) \Id \gamma (z)^{-1}=\Id$.
\end{pf}

\begin{proposition}
\label{periods}
Let $F_{1},\ldots,F_{\ell}$ be the  functions from
definition \ref{fndefn}.
Let $c_{1},\ldots, c_{\ell}$ be real constants.  Let $\left(\phi\left( t
\right),1 \right)$ denote any trajectory
of the Hamiltonian flow associated to
$F=\sum_{j=1}^\ell \ c_{j} F_{j}$.
If $\left(\phi(0),1\right)$ is a loop-regular element of the periodic Toda
phase space and $\phi(2\pi)=\phi(0)$ then
$c_{1}=c_{2}=\cdots=c_{\ell}=0$.  In other words, the period lattice for the
real flows of the Hamiltonians
$F_{1},\ldots,F_{\ell}$ is trivial.
\end{proposition}

\begin{pf}
By the factorization theorem \cite{RS} the Hamiltonian flow associated to
$F$ is given by
$$
\left( \phi(t),1 \right)=
\Adt^{*}_{b(t)}\left( \phi(0),1 \right),
$$
where
$\exp [t\grad F(\phi(0),1)]= k(t)^{-1}b(t)$
with $k(t)\in\kgph$ and $b(t)\in\bgph$.
If $\phi(2\pi) = \phi(0)$ then
$\left(\phi\left(0\right),1\right)
=\Adt^*_{b(2\pi)}\left(\phi\left(0\right),1\right)$, and
by lemma \ref{steph2} it follows that $b(2\pi)=\Id.$
Hence $\exp [2\pi \grad F (\phi(0),1)]\in\kgph$.

However $\grad F(\phi(0),1)\in i\kalh$ by corollary \ref{globalgrad}, and
so $\exp [2\pi \grad F (\phi(0),1)]$ belongs to $\kgph \cap \exp (i\kalh)
$.
Because of the Cartan decomposition
$\Lgpx = \kgph\cdot\exp(i\kalh)$
this forces $\exp [2\pi \grad F (\phi(0),1)] = \Id$,
and $\grad F (\phi(0),1) = 0$.
Hence, also by corollary \ref{globalgrad},
$c_{1}= c_{2}=\cdots = c_{\ell}=0$ as required.
\end{pf}

Notice that because of the decomposition $\Lgpx=\kgph\bgph$,
by the factorization theorem the flows induced on the Toda phase space by
the $F_j$'s are complete. Note also that because
${\rm M}_{(\phi,1)}\in\kgp$ for elements of the phase space, on the phase
space we have
$$
F_j (\phi,1) = \chi_j \left( {\rm M}_{(\phi,1)} \right)\; .
$$
\begin{corollary}
The functions $F_{1},\ldots,F_{\ell}$ constitute a
completely integrable system on the
Toda phase space.
\end{corollary}

\begin{pf}
It follows from the Kostant--Symes involution theorem that the functions
$F_{1},\ldots,F_{\ell}$
form a commuting family on the Toda phase space.
By proposition \ref{periods}, the Hamiltonian flows
of these functions on the Toda phase space are independent; hence the functions are independent.
Because the dimension of the Toda phase space is twice the number of functions, the system is completely integrable.
\end{pf}

\begin{corollary}
The functions $F_{1},\ldots,F_{\ell}$ do not all commute with the
generalized real symmetric periodic Toda lattice Hamiltonian.
\end{corollary}

\begin{pf}
Pick a loop-regular point $\left(\phi,1\right)$ in the phase space,
and set $\vec F =(F_1,\dots,F_{\ell})$.
Let $\vec F_c$ denote the connected 
component of the level set of $\vec F$
containing $(\phi,1)$ and note that 
this consists entirely of loop-regular elements. 
Because the flows induced by the $F_j$'s commute and are complete
there is an associated action of $(\real^{\ell},+)$ on $\vec F_c$.
Since $\hbox{dim}\vec F_c =\ell$ and 
the $F_j$'s are functionally independent on the 
level set the orbits of this action are open 
in $\vec F_c$.
But $\vec F_c$ is a disjoint union of such orbits and is connected,
hence $\vec F_c$ consists of a single orbit 
and the action is transitive. Because the period lattice is
trivial it follows that $\vec F_c$ is 
diffeomorphic to $\real^{\ell}$, and in particular is non-compact.

Now $\vec F_c$ is a closed subset of the Toda phase space, 
which itself can be regarded as a closed subset of $\real^{2\ell+1}$,
 cut out by the polynomial equation
$q_* \cdot \prod  q_i^{k_i} =1$.
Thus $\vec F_c$ is a closed non-compact subset of 
$\real^{2\ell+1}$, and consequently is unbounded.
But level sets of the Toda Hamiltonian
${\rm H}(\phi)=\kf{\phi}{\phi}$  are 
bounded and so the level set of $\rm H$ through $(\phi,1)$ cannot 
contain the level set $\vec F_c$. Hence it cannot contain
the flow of the $F_j$'s through $(\phi,1)$, 
and so $\rm H$ does not commute with all of the $F_j$'s.
\end{pf}

\newpage
% gloss.tex

\center{Glossary of Notation}

\begin{tabbing}
\hspace{2 cm} \= \kill         %tab format
$\al$  ..........\> a simple Lie algebra over $\cplx$ \\
$\hal$ ..........\> a Cartan subalgebra of $\al$ \\
$\bal$ ..........\> a Borel subalgebra of $\al$ \\
$\kal$ ..........\> a compact form of $\al$ \\
\\
$\gp$  ..........\> the simply connected Lie group corresponding
                                to $\al$ \\
$\csg$ ..........\> the Cartan subgroup of $\gp$ corresponding
                                to $\hal$ \\
$\bgp$ ..........\> the Borel subgroup of $\gp$ corresponding 
                                to $\bal$ \\
$\kgp$ ..........\> the compact subgroup of $\gp$ corresponding
                                to $\kal$ \\
\\
$\Lal$ ..........\> the loop algebra \\
$\balt$ ..........\> the algebra of loops with constant term in 
                                $\bal$, and no $z^{-1}$ terms \\
$\kalt$ ..........\> the algebra of loops with image in $\kal$\\
\\
$\Lgp$ ..........\> the loop group \\
$\bgpt$ ..........\> the subgroup of loops with constant term in 
                                $\bgp$, and no $z^{-1}$ terms \\
$\kgpt$ ..........\> the real subgroup of loops with image 
                                in $\kgp$\\
\\
$\Lalx$ ..........\> the centrally extended loop algebra \\
$\balh$ ..........\> the real subalgebra $\balt + \real$ \\
$\kalh$ ..........\> the real subalgebra $\kalt + i\real$ \\
\\
$\Lgpx$ ..........\> the centrally extended loop group \\
$\bgph$ ..........\> the real subgroup corresponding to $\balh$ \\
$\kgph$ ..........\> the real subgroup corresponding to $\kalh$
\end{tabbing}

\end{document}